\documentclass[aps,prb,floatfix,twocolumn,superscriptaddress]{revtex4-2}
\usepackage[utf8]{inputenc}
\usepackage{mathrsfs}
\usepackage{graphicx}
\usepackage{epstopdf}
\usepackage[english]{babel}
\usepackage{amsmath}
\usepackage{amssymb}
\usepackage{xcolor}
\usepackage{relsize}
\usepackage{CJK}
\usepackage{textcomp}
\usepackage{dsfont}
\usepackage{bm}
\usepackage{float}

\newcommand{\me}{\mathrm{e}}
\newcommand{\mi}{\mathrm{i}}

\newcommand{\dif}{\mathrm{d}}

\allowdisplaybreaks

\begin{document}

\title{Electrical-Circuit Simulation of the Uhlmann Phase}

\author{Yu-Huan Huang}
\affiliation{School of Physics, Southeast University, Jiulonghu Campus, Nanjing 211189, China}
\author{Yu Wang}
\affiliation{School of Physics, Southeast University, Jiulonghu Campus, Nanjing 211189, China}
\author{Jia-Chen Tang}
\affiliation{School of Physics, Southeast University, Jiulonghu Campus, Nanjing 211189, China}
\author{Xu-Yang Hou}
\affiliation{School of Physics, Southeast University, Jiulonghu Campus, Nanjing 211189, China}
\author{Hao Guo}
\email{guohao.ph@seu.edu.cn}
\affiliation{School of Physics, Southeast University, Jiulonghu Campus, Nanjing 211189, China}
\affiliation{Hefei National Laboratory, University of Science and Technology of China, Hefei 230088, China}
\begin{abstract}
The Uhlmann phase extends the concept of geometric phases to mixed quantum states through a parallel-transport condition on purification amplitudes, but its experimental realization has so far required sophisticated quantum platforms with carefully engineered auxiliary degrees of freedom. In this work, we reformulate the Uhlmann parallel-transport condition as a linear matrix differential equation and vectorize it to obtain an effective dynamical generator. This generator can be directly mapped onto the admittance matrix of a classical RC circuit, thereby translating the Uhlmann dynamics into the evolution of circuit node voltages. We illustrate the mapping using the equatorial-loop model and, via a rotating-frame transformation followed by a real decomposition, derive a time-independent, real-valued dynamical system suitable for analog implementation. LTspice simulations of the resulting active RC network faithfully reproduce the Uhlmann geometric phase and its topological transition at the critical purity, demonstrating that classical electrical circuits offer a simple and accessible platform for probing mixed-state geometric phases.
\end{abstract}

\maketitle
\section{Introduction}
Geometric phases are fundamental to the topological characterization of quantum systems~\cite{Bohm_GPbook,ChruscinskiBook}.
The Berry phase, acquired during adiabatic cyclic evolution of pure states, underpins the modern classification of topological insulators and superconductors through the Berry curvature and related invariants~\cite{Berry1984,TKNN,Haldane,KaneRMP,ZhangSCRMP,KaneMele,KaneMele2,ChiuRMP,Bernevigbook,BernevigPRL,MoorePRB,FuLPRL,Bohm_GPbook,vanderbilt2018berry,cohen2019geometric}.
These phases have been observed in a variety of platforms, including superconducting qubits, ultracold atoms, and photonic systems~\cite{Leek2007,Atala2013,Wang2023}.
Nevertheless, the Berry-phase framework is inherently restricted to pure states and cannot directly describe mixed states arising from environmental coupling or thermal fluctuations.

For mixed states, a mathematically rigorous generalization is provided by the Uhlmann phase~\cite{Uhlmann1986,Uhlmann1992}, which is constructed based on the purification $W$ (the counterpart of pure-state wavefunctions) of density matrices via $\rho=WW^\dag$ (similar to $\rho=|\psi\rangle\langle\psi|$ for pure states) and a parallel-transport condition imposed on $W$.
This framework has become central to the study of topological phenomena at finite temperature and in open quantum systems~\cite{Huang_2014,Viyuela2015,PhysRevLett.112.130401,PhysRevLett.119.015702,PhysRevB.97.235141,PhysRevB.107.165415,prq8-c9ns}.
An interferometric formulation of mixed-state geometric phases was also developed by Sj\"{o}qvist \textit{et al.} and demonstrated in NMR experiments~\cite{PhysRevLett.85.2845,PhysRevLett.91.100403}.
Both approaches extend geometric phases beyond pure states, yet they rest on different physical principles.
Here we adopt the Uhlmann framework.
While the Berry phase can be understood via parallel transport of pure states, the Uhlmann parallel-transport condition acts in the enlarged purification space and is not automatically satisfied under standard Hamiltonian dynamics~\cite{PhysRevB.101.104310,PhysRevA.104.023303}, thus requiring auxiliary purification degrees of freedom and engineered evolution protocols.
The Uhlmann phase has been observed in advanced quantum platforms such as superconducting circuits, photonic quantum walks, and programmable quantum processors~\cite{Muller2012,Viyuela2018,mastandrea2025,wang2025}.
However, in those experiments the Uhlmann phase was extracted either from the purified state, an entangled state in the system-ancilla space, or directly from interferometric or tomographic measurements on the density matrix, without implementing the parallel transport of the purification itself.This leaves open the possibility of simulating the Uhlmann parallel-transport condition in an equivalent, transformed representation, from which the geometric phase can be extracted after post-processing.

A central obstacle to the direct physical realization of Uhlmann parallel transport is that the purification $W$ of a density matrix is generally a non-Hermitian matrix, whose dynamical implementation does not follow from standard Hamiltonian evolution.
In this work, we circumvent this difficulty by reformulating the Uhlmann parallel-transport condition as a linear evolution equation for $W$, and then applying a vectorization procedure that converts this matrix equation into an ordinary linear dynamical system.
The resulting vectorized equation is structurally identical to the Kirchhoff-law description of classical RC circuits, which provides the basis for simulating Uhlmann parallel transport with electrical networks.
This approach is motivated by the broader development of electrical-circuit platforms as a versatile tool for emulating topological and quantum-inspired phenomena, through the mapping between effective Hamiltonians and circuit admittance matrices~\cite{PhysRevX.5.021031,PhysRevLett.114.173902,Lee2018}.
Such platforms have successfully realized a variety of topological phases, including Chern insulators, non-Hermitian topological matter, synthetic gauge fields, and Floquet engineered dynamics~\cite{PhysRevLett.122.247702,Helbig2020,Chen2023,PhysRevResearch.6.023010,Zhang2025,Sun2024,Sahin2025Topolectrical}.

Following this strategy, we construct an effective dynamical generator that directly encodes the Uhlmann parallel-transport condition, and establish an explicit mapping between the resulting evolution equations and an active RC circuit network.
As a proof of principle, we simulate the cyclic evolution of a two-level mixed state using LTspice and faithfully reproduce the topological transition of the Uhlmann phase predicted by theory, demonstrating that classical electrical circuits provide a simple and experimentally accessible platform for probing mixed-state geometric phases.

The rest of the paper is organized as follows. Section~\ref{sec2} reformulates the Uhlmann parallel-transport condition as a linear matrix differential equation for the purification $W$,
vectorizes it into a standard linear dynamical system, and establishes the mapping to the admittance matrix of an active RC circuit. Section~\ref{sec3} specializes the general framework to the equatorial-loop model for a two-level system, constructs a real constant dynamical
generator through a permutation, a rotating-frame transformation, and a real-valued decomposition, presents the detailed active RC network implementation, and validates the approach with LTspice simulations that faithfully reproduce the Uhlmann geometric phase and its topological $\pi$-transition at the critical purity $r_c=\sqrt{3}/2$, including a Monte Carlo analysis of static resistor tolerances and their impact on the resolvability of the critical point. Section~\ref{Sec4} summarizes the results and discusses possible extensions.

\section{Mapping Uhlmann Parallel Transport onto Circuit Dynamics}\label{sec2}

For a full-rank density matrix $\rho$ of rank $N$, the purification is
introduced through $\rho = W W^{\dagger}$,
where $W$ is an $N\times N$ matrix.
Since $\rho$ is assumed full-rank, $W$ admits the unique polar-type
decomposition
$W = \sqrt{\rho}\,U$,
with $U$ an $N\times N$ unitary matrix that embodies the $U(N)$ gauge
freedom of the purification, generalizing the $U(1)$ phase of pure states.
When the density matrix traverses a smooth curve parametrized by $t$,
its purification $W(t)$ is said to satisfy the Uhlmann parallel-transport
condition if
\begin{equation}
W^{\dagger}\dot W = \dot W^{\dagger}W ,
\label{eq:parallel}
\end{equation}
where the dot denotes $\mathrm{d}/\mathrm{d}t$, and \(t\) is not necessarily physical time.
This condition minimizes the Hilbert--Schmidt distance between infinitesimally separated purifications, making the distance independent
of the gauge choice $U$, i.e., $U(N)$-invariant.

 We now introduce $\mathcal{A}_U = -\dif U\,U^\dag$, which is in fact the Uhlmann connection governing the evolution of $U$ along a transport path in the parameter space. Substituting the decomposition $W = \sqrt{\rho}\,U$ into Eq.~(\ref{eq:parallel}) and using this definition, one obtains the Sylvester equation for $\mathcal{A}_U$,
\begin{equation}
\rho\mathcal{A}_U + \mathcal{A}_U\rho = -[\dif \sqrt{\rho},\sqrt{\rho}] .
\label{eq:Sylvester}
\end{equation}
This equation can be solved explicitly by expanding $\rho$ in its eigenbasis,
$\rho = \sum_i \lambda_i |i\rangle\langle i|$ with $\lambda_i > 0$.
Projecting onto the eigenstates yields
\begin{equation}
\mathcal{A}_U = - \sum_{ij}
\frac{\langle i| [\mathrm{d}\sqrt{\rho}, \sqrt{\rho}] |j\rangle}
{\lambda_i + \lambda_j}
|i\rangle\langle j| ,
\label{eq:AU}
\end{equation}
where $\mathrm{d}$ acts along the transport path.
With $\mathcal{A}_U$ thus determined, the evolution of $U$ is governed by $\dot U = -\mathcal{A}_U(X)U$, where $\mathcal{A}_U(X)$ denotes the contraction of the Uhlmann connection with the tangent vector $X = \frac{\dif}{\dif t}$ of the transport path.
This further determines the evolution of the purification:
\begin{equation}
\dot W = \dot{\sqrt{\rho}}\,U + \sqrt{\rho}\,\dot U
= -Y_s W - W Y_a^{T} ,
\label{eq:W_dynamics}
\end{equation}
where we have introduced the matrices
\begin{equation}\label{YsYa}
Y_s = -\dot{\sqrt{\rho}} (\sqrt{\rho})^{-1}, \quad
Y_a^{T} = U^{\dagger} \mathcal{A}_U U .
\end{equation}

Here, the subscripts $s$ and $a$ denote the system and ancillary sectors,
respectively: $Y_s$ arises from the variation of $\sqrt{\rho}$ and acts on
the system index of $W$ from the left, while $Y_a$ describes the gauge
evolution in the ancillary space and acts from the right through $Y_a^T$.
With $\mathcal{A}_U$ determined by Eq.~(\ref{eq:AU}),
Eq.~(\ref{eq:W_dynamics}) ensures that $W(t)$ satisfies the
parallel-transport condition~(\ref{eq:parallel}).

For a cyclic evolution of duration $\tau$ with $\rho(0)=\rho(\tau)$,
the Uhlmann geometric phase is defined as
\begin{equation}
\Phi_U = \arg\,\mathrm{Tr}\bigl[W^{\dagger}(0)W(\tau)\bigr]=\arg\,\mathrm{Tr}\bigl[\rho(0)\mathcal{P}\me^{-\oint \mathcal{A}_U}\bigr] ,
\label{eq:Uphase}
\end{equation}
where $\mathcal{P}$ is the path ordering operator.

To map the matrix differential equation~(\ref{eq:W_dynamics}) onto a
classical circuit, we introduce a vectorized representation of the
purification,
\begin{equation}
V(t) = \mathrm{rvec}[W(t)]=\begin{pmatrix}
W_{00}(t) \\ W_{01}(t) \\ W_{10}(t) \\ W_{11}(t)
\end{pmatrix}.
\end{equation}
The row-wise vectorization satisfies the identity
$\mathrm{rvec}(A W B^{T}) = (A\otimes B)\,\mathrm{rvec}(W)$.
Applying this identity to Eq.~(\ref{eq:W_dynamics}) yields the linear
system
\begin{equation}
\frac{\mathrm{d}}{\mathrm{d}t} V(t) = -Y_{\mathrm{eff}}(t) V(t),
\label{eq:V_dynamics}
\end{equation}
with the effective dynamical generator
\begin{equation}
Y_{\mathrm{eff}}(t) = Y_s(t)\otimes I + I\otimes Y_a(t)
\label{eq:Yeff} .
\end{equation}
Because the row-wise vectorization preserves the Hilbert--Schmidt inner
product, we have
\begin{equation}
V^{\dagger}(0)V(\tau) = \mathrm{Tr}\bigl[W^{\dagger}(0)W(\tau)\bigr] .
\label{eq:overlap}
\end{equation}
Consequently, the Uhlmann phase can be directly extracted from the circuit
voltages without reconstructing $W$:
\begin{equation}\label{PhiVU}
\Phi_U = \arg\bigl[V^{\dagger}(0)V(\tau)\bigr] .
\end{equation}

Moreover, since the row-wise vectorization is invertible,
Eq.~(\ref{eq:V_dynamics}) is exactly equivalent to
Eq.~(\ref{eq:W_dynamics}): the vectorized dynamics enforces the full matrix
parallel-transport condition~(\ref{eq:parallel}), which comprises $N^2$ real
constraints, and not merely its scalar reduction~\cite{PhysRevA.104.023303}
$\mathrm{Tr}[W^{\dagger}\dot W]=0$, a necessary but not sufficient condition
obtained by tracing Eq.~(\ref{eq:parallel}) (see Appendix~\ref{app4}). This
distinction clarifies the physical content of the present implementation.
Implementations based on the purified state in the system-ancilla space
access only the single scalar constraint, so that the parallel transport of
the purification has not previously been implemented at the dynamical level;
to our knowledge, no previous quantum or classical experiment has realized
the full matrix condition. As the circuit evolves under the complete set of
constraints in an equivalent, transformed representation, it constitutes, to
our knowledge, the first equivalent physical realization of Uhlmann parallel
transport.

Equation~(\ref{eq:V_dynamics}) is structurally identical to the dynamics
of a linear RC circuit.  According to Kirchhoff's current law, such a
circuit is described by
\begin{equation}
C \frac{\mathrm{d}}{\mathrm{d}t} V(t) + G(t) V(t) = 0 ,
\end{equation}
where $C$ is the capacitance and $G(t)$ the admittance matrix.
After normalization, this becomes
\begin{equation}
\frac{\mathrm{d}}{\mathrm{d}t} V(t) = -Y(t) V(t),
\label{eq:circuit}
\end{equation}
with $Y(t) = C^{-1}G(t)$ the effective admittance matrix.
Comparing Eqs.~(\ref{eq:V_dynamics}) and (\ref{eq:circuit}) immediately
establishes the mapping $Y(t) = Y_{\mathrm{eff}}(t)$, thereby translating
the Uhlmann parallel-transport process into the evolution of circuit node
voltages.  Diagonal elements of $Y_{\mathrm{eff}}$ correspond to
admittances to ground, while off-diagonal elements describe inter-node
couplings; non-Hermitian terms can be implemented using active circuit
elements.
In this way, we obtain a closed, circuit-friendly formulation of the
Uhlmann geometric phase.

\section{Equatorial-Loop Model}\label{sec3}
\subsection{Effective Dynamical Generator}

To demonstrate the circuit mapping on a concrete and analytically
tractable example, we now specialize the general framework to a two-level
system undergoing cyclic evolution along the equator of the Bloch sphere.
The density matrix of a two-level system is parametrized as
\begin{equation}
\rho = \frac{1}{2}\bigl(I + r\,\boldsymbol{n}\cdot\boldsymbol{\sigma}\bigr),
\end{equation}
where $\boldsymbol{n}$ is the unit Bloch vector,
$\boldsymbol{\sigma}=(\sigma_x,\sigma_y,\sigma_z)^T$ the vector of Pauli
matrices, and $0\le r\le 1$ measures the mixedness.  We take a closed
equatorial trajectory,
\begin{equation}
\boldsymbol{n}=(\cos\phi,\sin\phi,0),\qquad \phi\in[0,2\pi],
\end{equation}
such that the density matrix becomes
$\rho(\phi)=\frac12\bigl(I+r\cos\phi\,\sigma_x+r\sin\phi\,\sigma_y\bigr)$.
Its square root can be written as $\sqrt{\rho}=a I + b\,\boldsymbol{n}\cdot\boldsymbol{\sigma}$
with the coefficients
\begin{align}\label{ab}
a,b=\frac12\left(\sqrt{\frac{1+r}{2}}\pm\sqrt{\frac{1-r}{2}}\right).
\end{align}
Introducing the parametrization $\phi=\phi(t)$ and the notation
$\dot\phi=\mathrm{d}\phi/\mathrm{d}t$, we define the auxiliary constants
\begin{equation}\label{ab2}
\alpha = 2\sqrt{1-r^2},\quad
\beta  = 1-\sqrt{1-r^2},
\end{equation}
which depend only on the mixedness parameter $r$ and remain constant along
the loop.  Using Eqs.~(\ref{eq:AU}) and (\ref{YsYa}), the building blocks $Y_s$ and $Y_a$ take the explicit forms
\begin{equation}
Y_s = -\mathrm{i}\dot\phi
\begin{pmatrix}
\frac{\beta}{\alpha} & -\frac{r}{\alpha}\,\mathrm{e}^{-\mathrm{i}\phi} \\[6pt]
\frac{r}{\alpha}\,\mathrm{e}^{\mathrm{i}\phi} & -\frac{\beta}{\alpha}
\end{pmatrix},
\quad
Y_a = \mathrm{i}\dot\phi
\begin{pmatrix}
\frac{\beta}{2} & 0 \\[4pt]
0 & -\frac{\beta}{2}
\end{pmatrix}.
\end{equation}
Substituting these into the effective dynamical generator
$Y_{\mathrm{eff}} = Y_s\otimes I + I\otimes Y_a$ yields
\begin{align}\label{Yeff}
&Y_{\mathrm{eff}}(t)\notag\\=&
\mathrm{i}\begin{pmatrix}
\dot\phi\bigl(\frac{\beta}{2}-\frac{\beta}{\alpha}\bigr) & 0 &
\dot\phi\frac{r}{\alpha}\,\mathrm{e}^{-\mathrm{i}\phi} & 0 \\[6pt]
0 & -\dot\phi\bigl(\frac{\beta}{2}+\frac{\beta}{\alpha}\bigr) & 0 &
\dot\phi\frac{r}{\alpha}\,\mathrm{e}^{-\mathrm{i}\phi} \\[6pt]
-\dot\phi\frac{r}{\alpha}\,\mathrm{e}^{\mathrm{i}\phi} & 0 &
\dot\phi\bigl(\frac{\beta}{2}+\frac{\beta}{\alpha}\bigr) & 0 \\[6pt]
0 & -\dot\phi\frac{r}{\alpha}\,\mathrm{e}^{\mathrm{i}\phi} & 0 &
\dot\phi\bigl(\frac{\beta}{\alpha}-\frac{\beta}{2}\bigr)
\end{pmatrix}.
\end{align}

For any monotonic parametrization $\phi=\phi(t)$, the chain rule gives
$\frac{\mathrm{d}V}{\mathrm{d}t}
=\dot{\phi}(t)\frac{\mathrm{d}V}{\mathrm{d}\phi}
=-\dot{\phi}(t)Y_{\rm eff}(\phi)V$, so that
Eq.~(\ref{eq:V_dynamics}) becomes
\begin{align}
\frac{\mathrm{d}V}{\mathrm{d}\phi}
=-Y_{\rm eff}(\phi)V .
\label{eq:eff}
\end{align}
Thus, the Uhlmann holonomy depends only on the geometric path, not on its
traversal speed. If the circuit is required to reproduce a prescribed
nonuniform $\phi(t)$, however, the dynamical matrix must be scaled by
$\dot{\phi}(t)$, requiring time-dependent resistances.
Extensions to non-equatorial paths are discussed in Appendix~\ref{app2}.

\subsection{Construction of a Real Constant Dynamical Generator}

The effective dynamical generator $Y_{\mathrm{eff}}(\phi)$ derived above
contains explicit phase factors $\mathrm{e}^{\pm\mathrm{i}\phi}$.
A direct implementation would require coupling admittances that vary
continuously with $\phi$, which significantly complicates the circuit
design. To obtain a time-independent dynamical matrix suitable for an
RC network, we proceed in three steps.

\subsubsection{Block-Diagonalization via Component Permutation}

First, we reorder the components of the vector $V(t)$ in
Eq.~(\ref{eq:V_dynamics}) to bring out the underlying block-diagonal
structure.  By inspecting the structure of $Y_{\mathrm{eff}}$, we adopt
the rearrangement
\begin{equation}
V(t)=\begin{pmatrix}
V_{00}(t) \\ V_{01}(t) \\ V_{10}(t) \\ V_{11}(t)
\end{pmatrix}
\;\xrightarrow{P}\;
\hat{V}(t)=\begin{pmatrix}
V_{00}(t) \\V_{10}(t) \\ V_{01}(t) \\ V_{11}(t)
\end{pmatrix},
\end{equation}
where $P$ is the permutation matrix
\begin{align}
P=\begin{pmatrix}
1&0&0&0\\
0&0&1&0\\
0&1&0&0\\
0&0&0&1
\end{pmatrix},
\end{align}
and $\hat{V}=PV$.
Under this reordering, $Y_{\mathrm{eff}}$ is transformed into an
effective Hamiltonian $H(\phi)$, and the evolution equation (\ref{eq:V_dynamics}) takes the
block-diagonal Schr\"odinger-like form
\begin{equation}
\frac{\mathrm{d}}{\mathrm{d}\phi}\hat{V}
=
-\mathrm{i}H(\phi)\hat{V},
\quad
H(\phi)=
\begin{pmatrix}
H_A(\phi)&0\\
0&H_B(\phi)
\end{pmatrix},
\label{eq:H}
\end{equation}
where the two blocks $H_A$ and $H_B$ respectively act on the subspaces
$(V_{00},V_{10})^T$ and $(V_{01},V_{11})^T$. Explicitly, the two blocks are given by
\begin{align}
H_A(\phi)&=
\begin{pmatrix}
\frac{\beta}{2}-\frac{\beta}{\alpha} & \frac{r}{\alpha}\mathrm{e}^{-\mathrm{i}\phi} \\[6pt]
-\frac{r}{\alpha}\mathrm{e}^{\mathrm{i}\phi} & \frac{\beta}{2}+\frac{\beta}{\alpha}
\end{pmatrix},\notag\\
H_B(\phi)&=
\begin{pmatrix}
-\frac{\beta}{2}-\frac{\beta}{\alpha} & \frac{r}{\alpha}\mathrm{e}^{-\mathrm{i}\phi} \\[6pt]
-\frac{r}{\alpha}\mathrm{e}^{\mathrm{i}\phi} & \frac{\beta}{\alpha}-\frac{\beta}{2}
\end{pmatrix}.
\end{align}
\subsubsection{Rotating-Frame Transformation and Phase Invariance}
The off-diagonal elements of $H_A$ and $H_B$ still contain
$\mathrm{e}^{\pm\mathrm{i}\phi}$.  In the second step we eliminate this
residual phase dependence by a rotating-frame transformation.  We introduce
\begin{equation}\label{Uphi}
\hat{U}(\phi)=
\begin{pmatrix}
\mathrm{e}^{-\mathrm{i}\phi/2} & 0 \\
0 & \mathrm{e}^{\mathrm{i}\phi/2}
\end{pmatrix},
\end{equation}
and write $\hat{V}(\phi) = \hat{U}(\phi)\tilde{V}(\phi)$, where
$\tilde{V}$ is the transformed state (which will directly correspond to
the circuit node voltages). Rewriting the Schrödinger-like equation in terms of \(\tilde V(\phi)\), we obtain
\begin{equation}
\tilde{H} = \hat{U}^{\dagger} H \hat{U} - \mathrm{i}\,\hat{U}^{\dagger}
\frac{\mathrm{d}\hat{U}}{\mathrm{d}\phi},
\label{eq:H_tilde}
\end{equation}
we obtain the parameter-independent effective Hamiltonians
\begin{align}
\tilde{H}_A &=
\begin{pmatrix}
\frac{\beta}{2}-\frac{\beta}{\alpha}-\frac{1}{2} & \frac{r}{\alpha} \\[6pt]
-\frac{r}{\alpha} & \frac{\beta}{\alpha}+\frac{\beta}{2}+\frac{1}{2}
\end{pmatrix},
\notag\\
\tilde{H}_B &=
\begin{pmatrix}
-\frac{\beta}{2}-\frac{\beta}{\alpha}-\frac{1}{2} & \frac{r}{\alpha} \\[6pt]
-\frac{r}{\alpha} & \frac{\beta}{\alpha}-\frac{\beta}{2}+\frac{1}{2}
\end{pmatrix}.
\label{eq:HAB}
\end{align}
All matrix elements are now real constants, making the system directly
realizable with a fixed RC network.

We emphasize that the permutation and rotating-frame transformation introduced above are not restricted to the equatorial loop and can be applied to other paths as well. The equatorial loop is special only in that these transformations reduce the dynamical generator to a parameter-independent matrix, thereby allowing its direct realization using a fixed-component circuit. For a general non-equatorial path, the same procedure remains applicable, but the resulting dynamical matrix generally retains an explicit dependence on the path parameter. When the path is traversed in physical time, this parameter dependence leads to a time-dependent circuit dynamical matrix whose realization requires time-varying circuit elements. A detailed derivation for constant-latitude loops is provided in Appendix~\ref{app2}. The implementation and validation of such time-dependent circuits are beyond the scope of the present work and are left for future investigation.

\subsubsection{Real-Valued Decomposition for Stable Implementation}

Although the rotating-frame transformation removes the explicit
$\phi$-dependence and preserves the Uhlmann phase transition, the
resulting evolution equation
$\frac{\mathrm{d}\tilde{V}}{\mathrm{d}t} = -\mathrm{i}\tilde{H}\tilde{V}$
still contains the imaginary unit multiplying the non-symmetric real
matrices $\tilde{H}_{A}$ and $\tilde{H}_{B}$, and such complex
coefficients cannot be implemented directly with real resistive
couplings.  We therefore rewrite the complex dynamics as an equivalent
real-valued system.

For the circuit realization considered below, we choose the uniform
parametrization $\phi(t)=\omega t$ and set $\omega=1$. As discussed
above, this choice fixes the circuit time scale but does not restrict
the geometric Uhlmann holonomy of the loop. Decomposing the complex state into its real and imaginary
parts,
\begin{equation}
\tilde{V} = \tilde{V}_{R} + \mathrm{i}\tilde{V}_{I},
\end{equation}
and substituting it into the Schr\"odinger-like equation yields the
coupled real equations
\begin{equation}
\frac{\mathrm{d}\tilde{V}_{R}}{\mathrm{d}t} = \tilde{H}\tilde{V}_{I},
\qquad
\frac{\mathrm{d}\tilde{V}_{I}}{\mathrm{d}t} = -\tilde{H}\tilde{V}_{R}.
\end{equation}
For the $A$ sector, this pair can be recast into a four-dimensional
real-valued system,
\begin{equation}
\frac{\mathrm{d}}{\mathrm{d}t}
\begin{pmatrix}
\tilde{V}_{R}^{A} \\[2pt]
\tilde{V}_{I}^{A}
\end{pmatrix}
=
\begin{pmatrix}
0 & \tilde{H}_{A} \\
-\tilde{H}_{A} & 0
\end{pmatrix}
\begin{pmatrix}
\tilde{V}_{R}^{A} \\[2pt]
\tilde{V}_{I}^{A}
\end{pmatrix}.
\end{equation}
Since the $A$ and $B$ sectors remain decoupled, they can be assembled into
an $8\times8$ real-valued dynamical matrix.  Defining the global voltage
vector as
\begin{equation}
V = \bigl(
\tilde{V}_{R}^{A},\,
\tilde{V}_{I}^{A},\,
\tilde{V}_{R}^{B},\,
\tilde{V}_{I}^{B}
\bigr)^{\mathsf{T}},
\end{equation}
the final evolution equation takes the form
$\frac{\mathrm{d}V}{\mathrm{d}t} = M_{8} V$, with
\begin{equation}\label{M8}
M_{8} =
\begin{pmatrix}
0 & \tilde{H}_{A} & 0 & 0 \\
-\tilde{H}_{A} & 0 & 0 & 0 \\
0 & 0 & 0 & \tilde{H}_{B} \\
0 & 0 & -\tilde{H}_{B} & 0
\end{pmatrix}.
\end{equation}

The resulting dynamical matrix $M_8$ is purely real and block diagonal,
with one $4\times4$ block per sector.  The eigenvalues of
$\tilde{H}_{A,B}$ are real, $\pm\frac{\beta}{2}\pm\frac12$ (with
independent signs), so the spectrum of each $4\times4$ block is purely
imaginary, $\pm\mi\bigl(\frac{\beta}{2}\pm\frac12\bigr)$, and the
dynamics is bounded and oscillatory.  The real decomposition preserves
this spectrum, so the circuit inherits the oscillatory character of the
complex system, yielding a stable linear system that is directly
amenable to implementation using a classical RC circuit.

\subsection{Active RC Network Implementation of the Dynamical Generator}
\label{sec:circuit_implementation}

\subsubsection{Circuit Architecture and Design Principles}

The real-valued dynamical equation derived in the previous section,
$\frac{\mathrm{d}V}{\mathrm{d}t} = M_8 V$, can be directly mapped onto an
eight-node active RC network.  Each state variable is assigned to a
physical voltage node, and the couplings prescribed by the matrix elements
$M_{ij}$ are realized through appropriate resistive connections.

We emphasize that the circuit node voltages represent the transformed
dynamical variables, not the matrix elements of the original purification
amplitude $W$. The Uhlmann phase is recovered from these voltages through
the post-processing procedure described below.

Every node is built around a multi-input inverting integrator: its non-inverting input is grounded, and the inverting input serves as a virtual-ground summation node.  The output voltage $V_i$ is fed back through a capacitor $C$, while all coupled node voltages $V_j$ are
connected to the summation node through resistors $R_{ij}$.  Kirchhoff's
current law at the summation node yields
\begin{equation}
C\frac{\mathrm{d}V_i}{\mathrm{d}t} = -\sum_j \frac{V_j}{R_{ij}} .
\end{equation}
Comparing this with the target equation
$\frac{\mathrm{d}V_i}{\mathrm{d}t} = \sum_j M_{ij} V_j$ immediately gives
the component-matrix correspondence
\begin{equation}
M_{ij} = -\frac{1}{R_{ij}C}.
\label{eq:resistor_mapping}
\end{equation}
A uniform feedback capacitance $C = 1\,\mu\mathrm{F}$ is adopted
throughout to set the overall timescale.

The sign of each coupling is implemented according to the matrix element
$M_{ij}$.  Negative couplings are realized by a direct resistive
connection from the source node to the target integrator, because the
inverting integrator intrinsically introduces a minus sign.  Positive
couplings require an additional sign reversal, which is accomplished by
inserting a unity-gain inverting amplifier before the integrator input.
Thus the entire network is constructed from only two elementary building
blocks: multi-input inverting integrators and unity-gain inverting
amplifiers, interconnected by appropriately chosen resistors.

\subsubsection{Explicit Node Equations and Component Values}

The dynamics naturally decouples into two independent sectors, labeled
$A$ and $B$.  Taking the $A$ block as an example, we expand the state
vector in the basis
$(\tilde{V}_R^{A0},\tilde{V}_R^{A1},\tilde{V}_I^{A0},\tilde{V}_I^{A1})^{T}$.
The corresponding real-valued dynamical matrix reads
\begin{equation}
M_A =
\begin{pmatrix}
0 & 0 & h_{00} & h_{01}\\
0 & 0 & h_{10} & h_{11}\\
-h_{00} & -h_{01} & 0 & 0\\
-h_{10} & -h_{11} & 0 & 0
\end{pmatrix},
\end{equation}
with $h_{00}=\frac{\beta}{2}-\frac{\beta}{\alpha}-\frac12$, $h_{01}=\frac{r}{\alpha}$, $h_{10}=-\frac{r}{\alpha}$ and $h_{11}=\frac{\beta}{2}+\frac{\beta}{\alpha}+\frac12$.
Expressing the node equations explicitly,
\begin{align}
\frac{\mathrm{d}}{\mathrm{d}t}\tilde{V}^{A0}_R
&= h_{00}\tilde{V}^{A0}_I + \frac{r}{\alpha}\tilde{V}^{A1}_I,\notag\\[2pt]
\frac{\mathrm{d}}{\mathrm{d}t}\tilde{V}^{A1}_R
&= h_{11}\tilde{V}^{A1}_I - \frac{r}{\alpha}\tilde{V}^{A0}_I,\notag\\[2pt]
\frac{\mathrm{d}}{\mathrm{d}t}\tilde{V}^{A0}_I
&= -h_{00}\tilde{V}^{A0}_R - \frac{r}{\alpha}\tilde{V}^{A1}_R,\notag\\[2pt]
\frac{\mathrm{d}}{\mathrm{d}t}\tilde{V}^{A1}_I
&= -h_{11}\tilde{V}^{A1}_R + \frac{r}{\alpha}\tilde{V}^{A0}_R,
\end{align}
which uniquely determine the circuit topology.  Negative coefficients
correspond to direct resistive couplings, whereas positive coefficients
require an additional unity-gain inverter.  Since the signs of $h_{00}$,
$h_{11}$, and $r/\alpha$ remain unchanged over the whole parameter range,
varying the purity parameter $r$ only modifies the resistance values
without altering the circuit topology.

For the $A$ block, only three types of coupling resistors are needed:
\begin{equation}
R_{A0}=\frac{1}{|h_{00}|C},\quad
R_{\kappa}=\frac{\alpha}{rC},\quad
R_{A1}=\frac{1}{|h_{11}|C}.
\end{equation}
The $B$ block is constructed analogously, with the matrix entries of
$\tilde{H}_A$ replaced by the corresponding entries of $\tilde{H}_B$
appearing in the $B$ block of $M_8$ in Eq.~(\ref{M8}).  The complete
eight-node system therefore requires eight unity-gain inverters, four in
each block.

Although the vectorized formulation is formally applicable to
higher-dimensional systems, this formal generality does not imply a
practically scalable circuit implementation. For an $N$-level system, the
$N^2$ complex components of the purification amplitude require at least
$2N^2$ real voltage nodes. A fully connected implementation then requires
up to $O(N^4)$ coupling resistors, and the number of auxiliary operational
amplifiers grows accordingly, posing a practical complexity bottleneck for
higher-dimensional circuit implementations.
\begin{figure}[ht]
  \centering
  \includegraphics[width=0.48\textwidth]{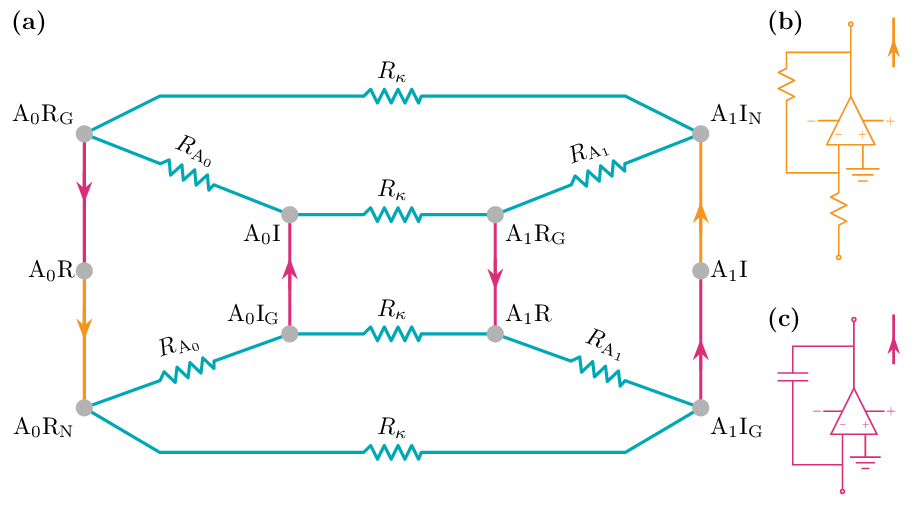}\\
  \caption{Circuit implementation of block $A$. (a) Circuit implementing the dynamics of the state variables $\tilde{V}_R^{A0}$, $\tilde{V}_I^{A0}$, $\tilde{V}_R^{A1}$, and $\tilde{V}_I^{A1}$. (b) Inverting amplifier with a gain of $-1$ used in the positive-coefficient branches corresponding to $\tilde{V}_R^{A0}$ and $\tilde{V}_I^{A1}$. (c) Inverting integrator. All integrators use $1\,\mu\mathrm{F}$ capacitors, and all operational amplifiers are AD823A.}
  \label{fig:A_ele}
\end{figure}

\subsubsection{Initialization and Measurement Protocol}

The initial state of the circuit is prepared by mapping the theoretical
purification amplitude $W(0)=\sqrt{\rho(0)}$ onto the integrator voltages.
At $\phi=0$, $\sqrt{\rho(0)} = \begin{pmatrix} a & b \\ b & a \end{pmatrix}$
with $a,b=\frac12\left(\sqrt{\frac{1+r}{2}}\pm\sqrt{\frac{1-r}{2}}\right)$.
To express the dimensionless initial amplitudes in a convenient
voltage scale, we introduce a global scaling factor
$\mathrm{amp}=10$. Since the circuit equations are linear and
homogeneous, this factor rescales all node voltages by
$\mathrm{amp}$ and the unnormalized overlap by
$\mathrm{amp}^2$, but does not affect the normalized overlap or
the extracted Uhlmann phase.

Set $\upsilon_a = a\,\mathrm{amp}$, $\upsilon_b = b\,\mathrm{amp}$.
The initial voltages of the eight integrator nodes are then
\begin{equation}
V(0) =
\bigl(\upsilon_a,\upsilon_b,0,0,\;
\upsilon_b,\upsilon_a,0,0\bigr)^{T},
\end{equation}
whose components correspond to
\begin{equation}
(V_R^{A0},V_R^{A1},V_I^{A0},V_I^{A1},\;
V_R^{B0},V_R^{B1},V_I^{B0},V_I^{B1})^{T}.
\end{equation}
Representative resistance values and initial voltages for several values
of $r$ are listed in Table~\ref{tab:parameters}.

\begin{table}[htbp]
    \centering
    \caption{Coupling resistances and initial voltage settings for
             different values of the purity parameter $r$. All resistance values are in $k\Omega$ (with $C=1\,\mu\mathrm{F}$).}
    \begin{tabular}{cccccccc}
    \hline
    \normalsize
    $r$ & $R_{A0}$ & $R_{\kappa}$ & $R_{A1}$ &
    $R_{B0}$ & $R_{B1}$ &
    $a\,\mathrm{amp}$ & $b\,\mathrm{amp}$ \\
    \hline
    0.300 & 1996 & 6360 & 1828 & 1828 & 1996 & 6.989 & 1.073 \\
    0.500 & 1959 & 3464 & 1552 & 1552 & 1959 & 6.830 & 1.830 \\
    0.700 & 1795 & 2040 & 1186 & 1186 & 1795 & 6.546 & 2.673 \\
    0.866 & 1333 & 1155 & 800  & 800  & 1333 & 6.124 & 3.535 \\
    0.900 & 1156 & 969  & 700  & 700  & 1156 & 5.991 & 3.755 \\
    \hline
    \end{tabular}
    \label{tab:parameters}
\end{table}

As an illustrative example, Fig.~\ref{fig:A_ele} shows the circuit
implementation of block $A$; for different values of the purity parameter
$r$, the configuration remains unchanged, and only the coupling resistances
and initial voltage amplitudes are adjusted according to
Table~\ref{tab:parameters}. The $B$ block has an identical structure and
is omitted for clarity.

For each value of $r$, the resistance values and initial voltages are
determined from the effective dynamical matrix, and transient simulations
are performed in LTspice over one complete evolution cycle
$\phi:0\rightarrow2\pi$.  The circuit dynamics are solved by LTspice,
while parameter sweeping, data extraction, and phase calculations are
handled automatically through Python scripts.

After the simulation, the final voltages of the eight nodes are used to
reconstruct the complex voltage vector
\begin{equation}
|V_c(T)\rangle =
\begin{pmatrix}
V_R^{A0}(T)+\mathrm{i} V_I^{A0}(T)\\
V_R^{A1}(T)+\mathrm{i} V_I^{A1}(T)\\
V_R^{B0}(T)+\mathrm{i} V_I^{B0}(T)\\
V_R^{B1}(T)+\mathrm{i} V_I^{B1}(T)
\end{pmatrix}.
\end{equation}

The measured overlap is $I_{\exp} = -\langle V_c(0)|V_c(T)\rangle$.
When the global scaling factor is taken into account, the normalized overlap is
$I = I_{\exp}/\mathrm{amp}^{2}$, its magnitude is
$|I| = \sqrt{(\mathrm{Re}I_{\exp})^{2}+(\mathrm{Im}I_{\exp})^{2}}/\mathrm{amp}^{2}$,
and the Uhlmann geometric phase is obtained as
$\Phi_U = \arg(I) = \mathrm{atan2}(\mathrm{Im}I_{\exp},\mathrm{Re}I_{\exp})$.
This procedure establishes a direct experimental pathway from the measured
circuit voltages to the Uhlmann geometric phase.

\subsection{LTspice Simulation Results and Comparison with Theory}
\label{sec3D}

\begin{figure}[ht]
  \centering
  \includegraphics[width=3.4in]{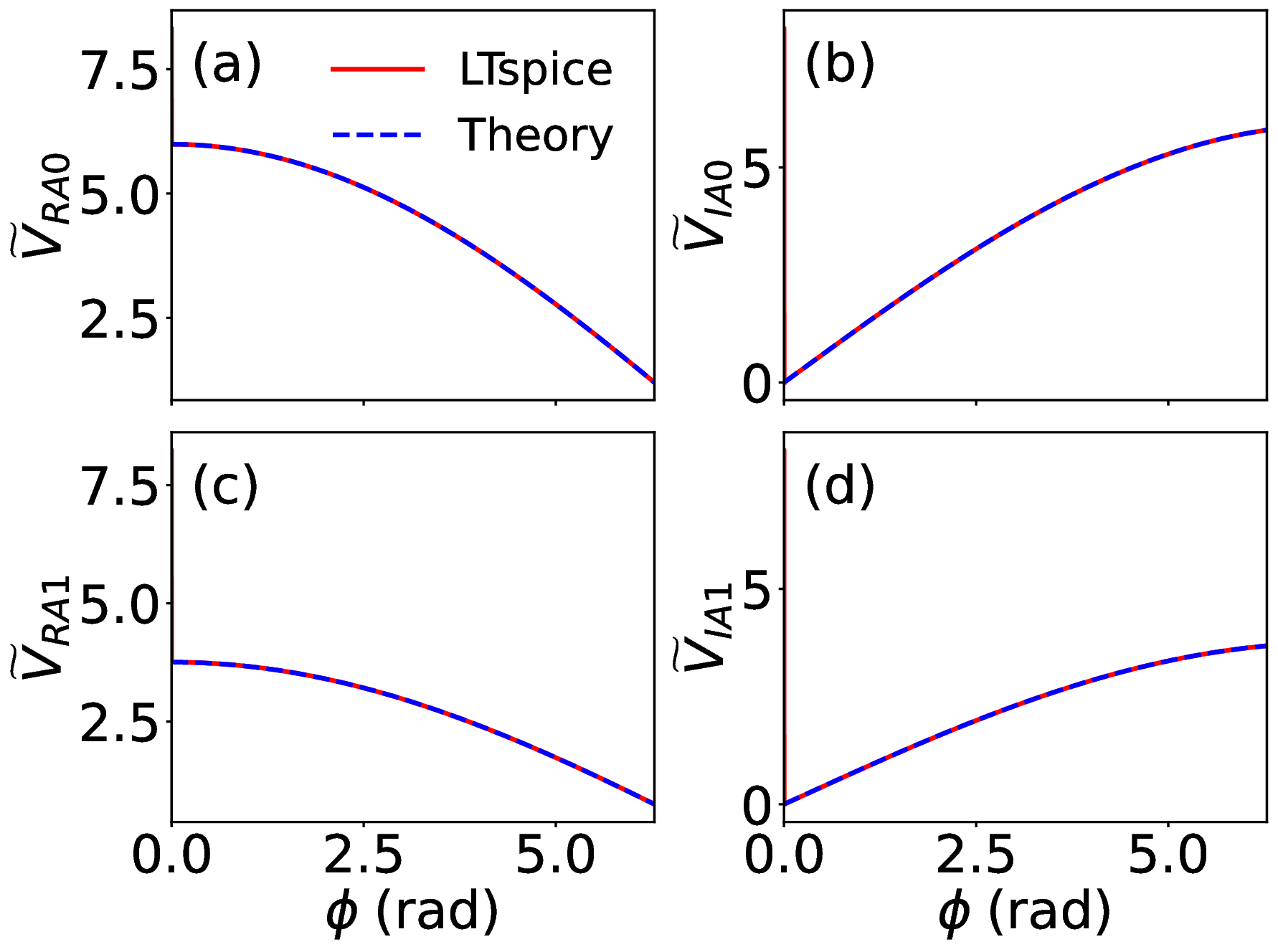}
  \caption{Comparison between LTspice simulations and theoretical
           predictions for the four state variables in block $A$ at
           $r=0.9$. Red solid lines: circuit simulation; blue dashed
           lines: theory.}
  \label{fig:A}
\end{figure}

To validate the proposed circuit, we first examine a representative case
deep in the non-trivial Uhlmann phase regime.  Figure~\ref{fig:A} shows
the time evolution of the four state variables of block $A$ at
$r=0.9$: $\tilde{V}_R^{A0}$, $\tilde{V}_I^{A0}$, $\tilde{V}_R^{A1}$, and
$\tilde{V}_I^{A1}$.  The LTspice simulations (red solid lines) are
compared directly with the exact theoretical solutions (blue dashed lines).

After a brief initial transient (about $1\,\mu\mathrm{s}$) caused by the
internal dynamics of the AD823A operational-amplifier macro-model, the
circuit voltages quickly lock onto the prescribed initial condition and
track the theoretical trajectories with high fidelity over the entire
cycle. This close agreement confirms that the active RC
network faithfully implements the transformed real-valued dynamics
constructed above, from which the Uhlmann phase is extracted after
post-processing.

\begin{figure}[ht]
    \centering
    \includegraphics[width=3.4in]{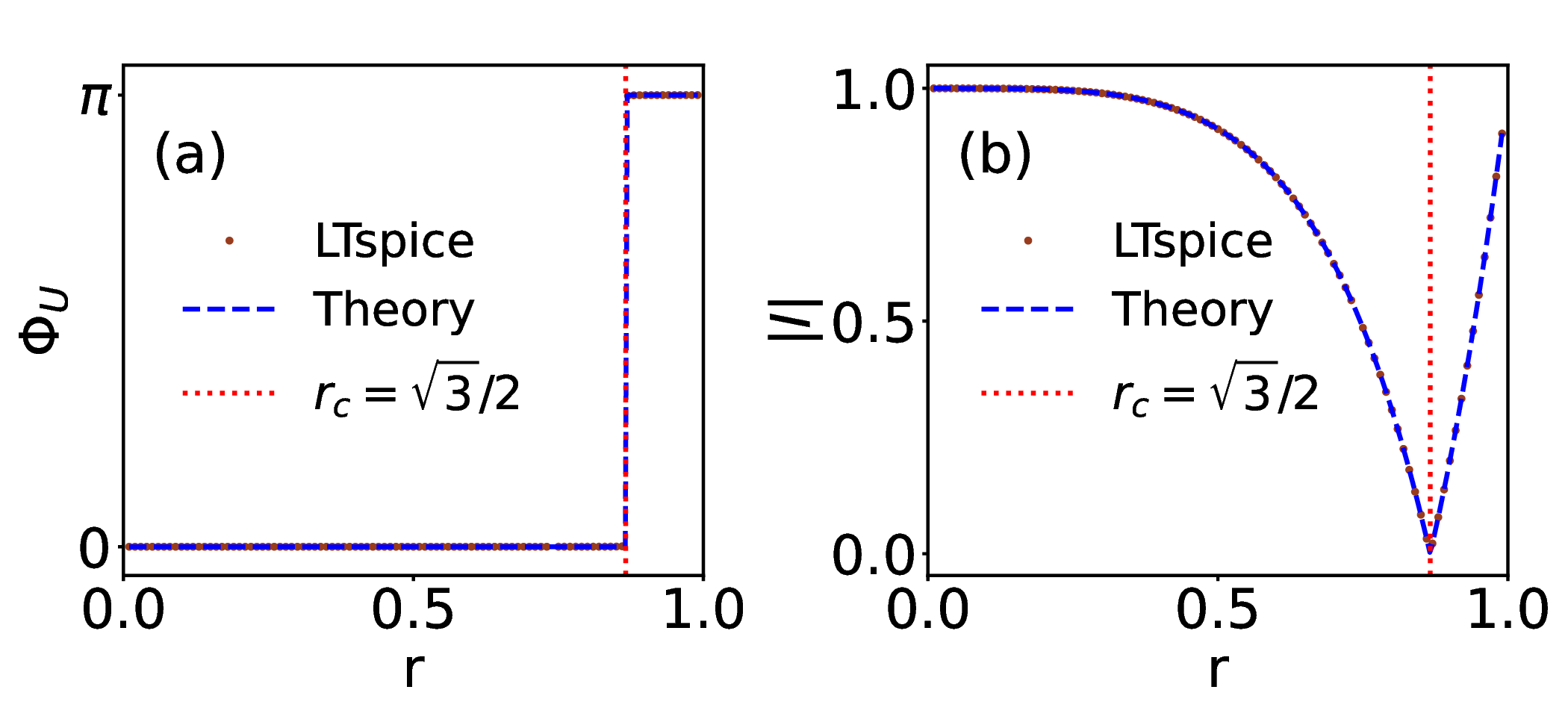}
    \caption{(a) Uhlmann geometric phase $\Phi_U$ as a function of $r$.
             (b) Normalized overlap magnitude $|I|$ as a function of $r$. Circles denote LTspice simulation results; dashed lines are the theoretical predictions}
    \label{fig:phase_results}
\end{figure}

A more stringent test is provided by the global geometric quantities.
Figure~\ref{fig:phase_results} summarizes the Uhlmann phase $\Phi_U$ and
the normalized overlap magnitude $|I|$ as functions of the purity
parameter $r$, sweeping across the topological transition.  The simulated
data points (circles) are superimposed on the theoretical curves (dashed
lines).

For $r < r_c = \sqrt{3}/2$, the Uhlmann phase stays at zero; for $r > r_c$, it jumps to $\pi$, exactly at the critical purity predicted by the analytical formula $\Phi_U = \arg[\cos(\pi\beta)]$ with $\beta = 1-\sqrt{1-r^{2}}$.  The overlap magnitude $|I|$ likewise
follows the theoretical expression $|I| = |\cos(\pi\beta)|$, exhibiting a pronounced dip near $r_c$ where the overlap vanishes. The simultaneous quantitative agreement of both quantities over the whole range of $r$ demonstrates that the classical circuit not only reproduces the local transformed dynamics but also captures the global topological signature of the Uhlmann phase.

\subsection{Static Resistor Tolerances and Critical-Point Resolvability}
\label{sec:tolerance}

To assess the robustness of the circuit implementation against component
imperfections, we perform Monte Carlo simulations of static resistor
tolerances in the $16$ coupling resistors of the complete circuit. For each
realization, the relative deviation of every resistor is independently
sampled from the uniform interval $[-\eta,\eta]$ and held fixed throughout
the scan over the purity parameter $r$, so that each realization represents
one fabricated nonideal circuit. We consider $\eta=0.5\%$ and $1.0\%$, with
$N=100$ realizations per tolerance level. The full Monte Carlo procedure and
the extraction of the effective critical purity are detailed in
Appendix~\ref{app3}.

\begin{figure}[t]
    \centering
    \includegraphics[width=3.4in]{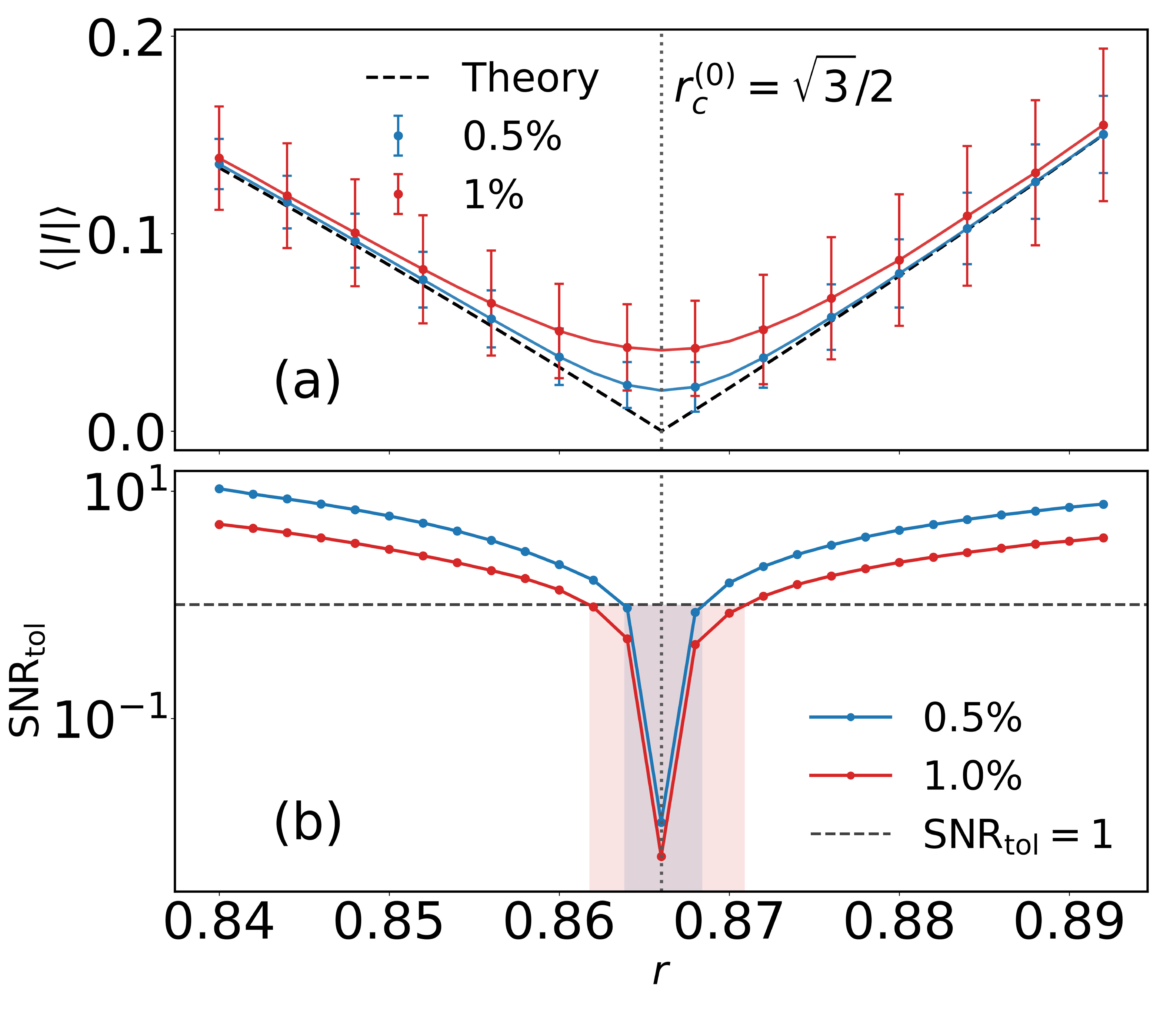}
    \caption{Monte Carlo results under static resistor tolerances.
    (a) Mean overlap amplitude $\langle |I| \rangle$ for resistor
    tolerances of $0.5\%$ and $1.0\%$; error bars denote the sample
    standard deviation $\sigma_{|I|}$. The black dashed curve is the
    ideal-theory result, and the gray dotted line marks
    $r_c^{(0)}=\sqrt{3}/2$.
    (b) Tolerance-limited signal-to-noise ratio
    $\mathrm{R}_{\mathrm{SNtol}}$. The horizontal dashed line marks
    $\mathrm{R}_{\mathrm{SNtol}}=1$, and the shaded regions indicate
    $\mathrm{R}_{\mathrm{SNtol}}\leq1$.}
    \label{fig:rc_tolerance}
\end{figure}

For each realization, the effective critical purity is extracted from the
local minimum of the overlap amplitude $|I(r)|$.
Fig.~\ref{fig:rc_tolerance} summarizes the overlap-amplitude statistics and
the resulting tolerance-limited resolvability near the ideal critical purity
$r_c^{(0)}=\sqrt{3}/2$.

As shown in Fig.~\ref{fig:rc_tolerance}(a), static resistor tolerances lift
and broaden the minimum of the mean overlap amplitude near $r_c^{(0)}$: the
minima of different realizations occur at slightly different purity values
and, since $|I|$ is non-negative, averaging over the ensemble produces a
nonzero, broadened valley. The extracted critical purities are
\begin{align}
r_c^{\mathrm{MC}}
&=
0.8662\pm0.0029,
\qquad \eta=0.5\%,
\notag\\
r_c^{\mathrm{MC}}
&=
0.8661\pm0.0058,
\qquad \eta=1.0\%.
\label{eq:result_one_rc}
\end{align}
Here, the values following $\pm$ are the sample standard deviations over
circuit realizations, not standard errors of the mean. The mean values
remain consistent with $r_c^{(0)}$ within the present sampling precision,
whereas doubling the tolerance doubles the spread, $\sigma_{r_c}\colon
0.0029\to0.0058$. Thus, the dominant effect of static resistor tolerances is
not a systematic displacement of the critical point, but a substantially
broader circuit-to-circuit distribution of the extracted critical purity.

To quantify how tolerances affect the resolvability of the overlap dip, we
define the tolerance-limited signal-to-noise ratio
\begin{align}
\mathrm{R}_{\mathrm{SNtol}}(r)
=
\frac{|I_0(r)|}{\sigma_{|I|}(r)},
\label{eq:snr_tol}
\end{align}
where $I_0(r)$ is the ideal overlap and $\sigma_{|I|}(r)$ is the sample
standard deviation of the overlap amplitude over the Monte Carlo ensemble.
We stress that $\mathrm{R}_{\mathrm{SNtol}}$ is an amplitude-resolvability
measure set by static tolerances, rather than the complete signal-to-noise
ratio of a physical circuit. Taking $\mathrm{R}_{\mathrm{SNtol}}=1$ as an
operational one-standard-deviation criterion and interpolating linearly
between adjacent scan points, we obtain the low-resolvability intervals
\begin{align}
r&\in[0.8638,\,0.8684],
\qquad \eta=0.5\%,
\notag\\
r&\in[0.8618,\,0.8709],
\qquad \eta=1.0\%.
\label{eq:snr_one_percent}
\end{align}
The interval width grows from $0.0046$ to $0.0091$, i.e., approximately in
proportion to $\eta$. Within these intervals, an amplitude-only measurement
cannot reliably localize the critical purity. The critical point itself is
not destroyed; rather, static resistor tolerances impose a finite
uncertainty window around its ideal position, within which the
overlap-amplitude minimum cannot be stably extracted.

Two remarks are in order. First, the topological $\pi$-jump can also be
detected through the sign change of $\operatorname{Re}I(r)$, which crosses
zero linearly at $r_c$, instead of through the minimum of $|I(r)|$. Such
zero-crossing detection is expected to be less sensitive to amplitude
fluctuations, and constitutes a natural noise-mitigation strategy, together
with the use of precision resistors, low-offset (auto-zero or chopper)
operational amplifiers, and averaging over repeated evolution cycles.
Second, the present analysis covers only static resistor deviations after
fabrication. Thermal noise, operational-amplifier noise, and offset drift
cannot be captured by fixed component perturbations; their treatment
requires a time-dependent stochastic extension of the circuit model with
parameters taken from a specific physical implementation. A quantitative
analysis of these effects will be pursued in future work.

\section{Conclusion}
\label{Sec4}

We have reformulated the Uhlmann parallel-transport condition as a linear
matrix differential equation and mapped it, via vectorization, onto the
admittance matrix of an active RC circuit. A permutation and a
rotating-frame transformation render the dynamical generator
time-independent, and a real-valued decomposition gives an equivalent
purely real linear system suitable for implementation with operational amplifiers, resistors, and capacitors. LTspice simulations of the
equatorial-loop model faithfully reproduce the Uhlmann geometric phase and
its $\pi$-transition at the critical purity $r_c=\sqrt3/2$, validating the
proposed mapping. Since the transformations mapping $W$ onto the
circuit variables are exact and invertible, the node-voltage evolution
implements the full matrix parallel-transport condition in an equivalent
representation, rather than only its scalar reduction
$\mathrm{Tr}[W^{\dagger}\dot W]=0$ accessible in purified-state
implementations.

Monte Carlo simulations further show that static resistor tolerances leave
the mean extracted critical purity consistent with the ideal value, while
the circuit-to-circuit spread grows with the tolerance, and the
tolerance-limited $\mathrm{R}_{\mathrm{SNtol}}$ defines a finite
low-resolvability interval around $r_c$ within which the overlap amplitude
alone cannot localize the transition. The significance of this approach lies
in offering a potentially accessible experimental route to simulating
Uhlmann parallel transport, rather than a computationally faster alternative
to numerical integration. Its limitations are the path-dependent parameters
required for general trajectories and the rapid growth of circuit elements
with system dimension; experimental validation in a physical active-RC
network, and extensions to time-dependent and higher-dimensional circuits,
remain open directions.

\section{Acknowledgments}
H. G. was supported by the Quantum Science and Technology-National Science and Technology Major Project (Grant No.2021ZD0301904) and the National Natural Science Foundation of China (Grant No. 12447216). X.-Y. H. was supported by the National Natural Science Foundation of China (Grant No.12405008).
\appendix
\section{One--Cycle Overlap and Phase Invariance}
\label{app1}

We show explicitly that neither the basis permutation $P$ nor the
rotating-frame transformation $\hat U(\phi)$ introduced in the main text
modifies the Uhlmann geometric phase.  The phase is defined by the overlap
of the original vectorized purification,
\begin{equation}
\Phi_U = \arg\!\bigl[V^\dagger(0)V(2\pi)\bigr] .
\end{equation}

\noindent\textbf{Step 1: permutation.}
Because $P$ is a unitary matrix ($P^\dagger P = I$), the overlap is
unchanged under the reordering $\hat V = P V$:
\begin{equation}
V^\dagger(0)V(2\pi) = \hat V^\dagger(0)\hat V(2\pi).
\end{equation}

\noindent\textbf{Step 2: rotating frame.}
The second transformation acts independently on each block,
$\hat V_j(\phi) = \hat U(\phi)\tilde V_j(\phi)$ with $j=A,B$, where $\hat U(\phi)$ is given by Eq.~(\ref{Uphi}).
This step removes the residual $\phi$-dependence from the Hamiltonians,
which become
\begin{equation}
\tilde H_A = \frac{\beta}{2}I + M , \quad
\tilde H_B = -\frac{\beta}{2}I + M ,
\end{equation}
with
\begin{equation}
M =
\begin{pmatrix}
-\frac{\beta}{\alpha}-\frac12 & \frac{r}{\alpha} \\[8pt]
-\frac{r}{\alpha} & \frac{\beta}{\alpha}+\frac12
\end{pmatrix}.
\end{equation}
Because $\tilde H_A$ and $\tilde H_B$ are constant matrices, the evolution operators
after one full cycle ($\phi$ from $0$ to $2\pi$) are simply
\begin{equation}
\tilde{\mathcal U}_{A,B}(2\pi) = \me^{-\mi 2\pi \tilde H_{A,B}} .
\end{equation}

Consider block $A$:
\begin{align}
\tilde{\mathcal U}_A(2\pi)
&= \me^{-\mi 2\pi(\frac{\beta}{2}I + M)}
   = \me^{-\mi\pi\beta}\,\me^{-\mi 2\pi M} .
\end{align}
Noting that $2\pi M = \pi(2M)$ and $(2M)^2 = I$, the second factor reduces to
\begin{equation}
\me^{-\mi\pi(2M)} = \cos(\pi)I - \mi\sin(\pi)(2M) = -I .
\end{equation}
Thus
\begin{equation}
\tilde{\mathcal U}_A(2\pi) = -\me^{-\mi\pi\beta} I .
\end{equation}
An identical calculation for block $B$ gives
\begin{equation}
\tilde{\mathcal U}_B(2\pi) = -\me^{\mi\pi\beta} I .
\end{equation}

The initial purification is $\sqrt{\rho} = \begin{pmatrix} a & b \\ b & a \end{pmatrix}$ with
$a,b = \frac12\bigl(\sqrt{\frac{1+r}{2}} \pm \sqrt{\frac{1-r}{2}}\bigr)$.
Its row-vectorization yields $V(0) = (a,b,b,a)^T$, which after permutation
decomposes as
\begin{equation}
\hat V_A(0) = \begin{pmatrix} a \\ b \end{pmatrix},\quad
\hat V_B(0) = \begin{pmatrix} b \\ a \end{pmatrix}.
\end{equation}
Using $V^\dagger V = \operatorname{Tr}(W^\dagger W) = \operatorname{Tr}(\rho) = 1$,
we have $2(a^2+b^2)=1$, hence $a^2+b^2 = 1/2$ and
\begin{equation}
\hat V_A^\dagger\hat V_A = \hat V_B^\dagger\hat V_B = \frac12 .
\end{equation}
Unitarity of $\hat U(\phi)$ preserves these norms,
\begin{equation}
\tilde V_A^\dagger\tilde V_A = \tilde V_B^\dagger\tilde V_B = \frac12 .
\end{equation}
The boundary values of $\hat U(\phi)$ are
$\hat U(0) = I$ and
$\hat U(2\pi) = \begin{pmatrix} \me^{-\mi\pi} & 0 \\ 0 & \me^{\mi\pi} \end{pmatrix} = -I$.
Consequently, at the end of the cycle,
\begin{equation}
\hat V_j(2\pi) = \hat U(2\pi)\tilde V_j(2\pi) = -\tilde V_j(2\pi) .
\end{equation}
We can now evaluate the overlap, inserting this sign together with the
evolution operators:
\begin{align}
&V^\dagger(0)V(2\pi)
= \hat V^\dagger(0)\hat V(2\pi) \notag\\
=& \sum_{j=A,B} \tilde V_j^\dagger(0)\,\bigl[-\tilde{\mathcal U}_j(2\pi)\bigr]\,   \tilde V_j(0) \notag\\
=& \tilde V_A^\dagger(0)\bigl[-(-\me^{-\mi\pi\beta}I)\bigr]\tilde V_A(0)
   + \tilde V_B^\dagger(0)\bigl[-(-\me^{\mi\pi\beta}I)\bigr]\tilde V_B(0) \notag\\
=& \cos(\pi\beta).
\end{align}
The two minus signs cancel, leaving exactly the overlap one obtains from
the original Uhlmann parallel-transport theory.  Therefore the Uhlmann phase
\begin{equation}
\Phi_U = \arg\!\bigl[\cos(\pi\beta)\bigr]
\end{equation}
is unchanged by the auxiliary transformations.  It exhibits the
topological $\pi$-jump at $\beta = 1/2$, i.e.\ at the critical purity
$r_c = \sqrt{3}/2$.

\section{Matrix and Scalar Forms of the Parallel-Transport Condition}
\label{app4}

For an $N$-dimensional system, the parallel-transport
condition~(\ref{eq:parallel}) requires the $N\times N$ matrix
$W^{\dagger}\dot W$ to be Hermitian, which amounts to $N^2$ real constraints
on the dynamics of the purification amplitude. A weaker, scalar condition is
obtained by tracing the matrix condition: Eq.~(\ref{eq:parallel}) implies
that $\mathrm{Tr}[W^{\dagger}\dot W]$ is real, while the normalization
$\mathrm{Tr}[\rho]=\mathrm{Tr}[W^{\dagger}W]=1$ forces its real part to
vanish, so that
\begin{equation}
\langle W|\tfrac{\dif}{\dif t}|W\rangle
=
\mathrm{Tr}\bigl[W^{\dagger}\dot W\bigr]=0 ,
\label{eq:scalar_pt}
\end{equation}
where $|W\rangle=\sum_{i}\sqrt{\lambda_{i}}\,|i\rangle\otimes U^{T}|i\rangle$
is the purified state associated with $W=\sqrt{\rho}\,U$, and the first
equality follows from the Hilbert--Schmidt isometry between purification
amplitudes and purified states.
Equation~(\ref{eq:scalar_pt}) is therefore a necessary but not sufficient
consequence of the matrix condition: it constitutes a single complex
equation whose real part is already fixed by normalization, leaving one
independent real constraint, in contrast to the $N^2$ real constraints of
Eq.~(\ref{eq:parallel}). For the two-level system considered in this work,
the matrix condition comprises four real constraints, only one of which is
captured by Eq.~(\ref{eq:scalar_pt}).

The construction of Sec.~\ref{sec2} maps the matrix condition itself, rather
than only its scalar reduction, onto the circuit dynamics. The row-wise
vectorization is an invertible linear map, so that
Eq.~(\ref{eq:V_dynamics}) is exactly equivalent to
Eq.~(\ref{eq:W_dynamics}), whose solution by construction satisfies
Eq.~(\ref{eq:parallel}). The permutation, rotating-frame transformation, and
real decomposition of Sec.~\ref{sec3} are exact reformulations that leave
the one-cycle overlap invariant, as shown in Appendix~\ref{app1}.
Consequently, the eight real node voltages of the circuit encode the four
complex components of $W$ in a transformed representation, and the circuit
evolution enforces all $N^2=4$ real constraints of the parallel-transport
condition at the level of the dynamical equations, up to the accuracy of the
circuit components and the numerical solver.

\section{Constant-Latitude Paths and Time-Dependent Generators}
\label{app2}

The fixed-component circuit in the main text is designed for the equatorial
loop. Here we follow the same procedure and extend the path to an arbitrary
constant-latitude loop. This shows that the method itself is not restricted
to the equator. For a general path, however, the resulting dynamical
matrix retains an explicit dependence on the path parameter.

We consider the closed path at a fixed polar angle $\theta_0$ (with
$s=\sin\theta_0$ and $c=\cos\theta_0$), $\phi\in[0,2\pi]$.
The corresponding density matrix is
$\rho(\phi)=\frac{1}{2}[I+rs(\sigma_x\cos\phi+\sigma_y\sin\phi)+rc\sigma_z]$,
where $0<r<1$. Its square root can be written as
\begin{align}
 \sqrt{\rho(\phi)}
 =aI+b
 \left[
 s\left(\sigma_x\cos\phi+\sigma_y\sin\phi\right)
 +c\sigma_z
 \right],
 \label{eq:app_sqrt_rho_latitude}
\end{align}
where $a$ and $b$ are given by Eq.~(\ref{ab}), and $\alpha$ and $\beta$
are defined in Eq.~(\ref{ab2}). These parameters satisfy
\begin{align}
 a^2-b^2=\frac{\alpha}{4},
 \quad
 ab=\frac{r}{4},
 \quad
 b^2=\frac{\beta}{4}.
 \label{eq:app_ab_relations}
\end{align}
Using $Y_s=-(\dif\sqrt{\rho}/\dif\phi)(\sqrt{\rho})^{-1}$ from the main
text, we obtain
\begin{align}
 Y_s(\phi)
 ={}&-\frac{rs}{\alpha}
 \left(-\sigma_x\sin\phi+\sigma_y\cos\phi\right)
 \nonumber\\
 &+\mathrm{i}\frac{\beta sc}{\alpha}
 \left(\sigma_x\cos\phi+\sigma_y\sin\phi\right)
 -\mathrm{i}\frac{\beta s^2}{\alpha}\sigma_z.
 \label{eq:app_Ys_latitude_pauli}
\end{align}
From Eq.~(\ref{eq:app_sqrt_rho_latitude}), the Uhlmann connection along the
constant-latitude path is
\begin{align}
 A_U^\phi(\phi)
 =\frac{\mathrm{i}\beta}{2}
 \left[
 -sc\left(\sigma_x\cos\phi+\sigma_y\sin\phi\right)
 +s^2\sigma_z
 \right].
 \label{eq:app_AU_latitude_pauli}
\end{align}

Following the main text, the Uhlmann transporter and the ancillary-sector
generator satisfy
\begin{align}
 \frac{\mathrm{d}U}{\mathrm{d}\phi}
 &=-A_U^\phi(\phi)U(\phi),
 \quad U(0)=I,
 \nonumber\\
 Y_a^T(\phi)
 &=U^\dagger(\phi)A_U^\phi(\phi)U(\phi).
 \label{eq:app_U_equation}
\end{align}
Since $A_U^\phi$ is anti-Hermitian, Eq.~\eqref{eq:app_U_equation} gives
\begin{align}
 \frac{\mathrm{d}U^\dagger}{\mathrm{d}\phi}
 =U^\dagger A_U^\phi.
 \label{eq:app_dUdagger}
\end{align}
Differentiating $Y_a^T$ directly, we find
\begin{align}
 \frac{\mathrm{d}Y_a^T}{\mathrm{d}\phi}
 &={}
 \frac{\mathrm{d}U^\dagger}{\mathrm{d}\phi}A_U^\phi U
 +U^\dagger\frac{\mathrm{d}A_U^\phi}{\mathrm{d}\phi}U
 +U^\dagger A_U^\phi\frac{\mathrm{d}U}{\mathrm{d}\phi}
 \nonumber\\
 &=U^\dagger A_U^\phi A_U^\phi U
 +U^\dagger\frac{\mathrm{d}A_U^\phi}{\mathrm{d}\phi}U
 -U^\dagger A_U^\phi A_U^\phi U
 \nonumber\\
 &=U^\dagger(\phi)
 \frac{\mathrm{d}A_U^\phi(\phi)}{\mathrm{d}\phi}
 U(\phi).
 \label{eq:app_dYa}
\end{align}
Equation~\eqref{eq:app_AU_latitude_pauli} gives
\begin{align}
 \frac{\mathrm{d}A_U^\phi}{\mathrm{d}\phi}
 =\frac{\mathrm{i}\beta sc}{2}
 \left(\sigma_x\sin\phi-\sigma_y\cos\phi\right).
 \label{eq:app_dAU}
\end{align}
For a generic non-equatorial constant-latitude loop, $sc\neq0$, and hence $\frac{\mathrm{d}A_U^\phi}{\mathrm{d}\phi}\neq0$, $\frac{\mathrm{d}Y_a^T}{\mathrm{d}\phi}\neq0$.
Thus $Y_a(\phi)$ generally depends on the path parameter. This differs from
the equatorial path, where $s=1$ and $c=0$, so that $A_U^\phi=\frac{\mathrm{i}\beta}{2}\sigma_z$, $Y_a^T=\frac{\mathrm{i}\beta}{2}\sigma_z$,
and both matrices are constant.

We now apply the row-vectorization procedure of the main text, which gives
\begin{align}
 \frac{\mathrm{d}V}{\mathrm{d}\phi}
 =-Y_{\mathrm{eff}}(\phi)V
\end{align}
with $Y_{\mathrm{eff}}(\phi)
 =Y_s(\phi)\otimes I+I\otimes Y_a(\phi)$.
Writing the two $2\times2$ matrices as
\begin{align}
 Y_s=
 \begin{pmatrix}
 p_{00}&p_{01}\\
 p_{10}&p_{11}
 \end{pmatrix},
 \quad
 Y_a=
 \begin{pmatrix}
 q_{00}&q_{01}\\
 q_{10}&q_{11}
 \end{pmatrix},
 \label{eq:app_matrix_elements}
\end{align}
and applying the same permutation $P$ as in the main text, we obtain the
permuted generator
\begin{align}
 \hat Y_{\mathrm{eff}}
 &=P Y_{\mathrm{eff}}P^\dagger
 \nonumber\\
 &=
 \begin{pmatrix}
 p_{00}+q_{00}&p_{01}&q_{01}&0\\
 p_{10}&p_{11}+q_{00}&0&q_{01}\\
 q_{10}&0&p_{00}+q_{11}&p_{01}\\
 0&q_{10}&p_{10}&p_{11}+q_{11}
 \end{pmatrix}
 \nonumber\\
 &=
 \begin{pmatrix}
 Y_s+q_{00}I&q_{01}I\\
 q_{10}I&Y_s+q_{11}I
 \end{pmatrix}.
 \label{eq:app_permuted_Yeff}
\end{align}
For the equatorial path, $Y_a$ is diagonal and constant, so that
$q_{01}=q_{10}=0$, and Eq.~\eqref{eq:app_permuted_Yeff} separates into the
two independent $A$ and $B$ sectors of the main text. For a generic
non-equatorial constant-latitude loop, $Y_a(\phi)$ varies with $\phi$ and
generally has nonzero off-diagonal elements; the two sectors are therefore
coupled.

We next apply the rotating-frame transformation $\hat U(\phi)$ of
Eq.~(\ref{Uphi}) to the permuted voltage vector, writing
\begin{align}
\hat V(\phi)
=
\left[I\otimes\hat U(\phi)\right]
\widetilde V(\phi),
\label{eq:app_transformed_voltage}
\end{align}
where $\widetilde V$ collects the rotating-frame variables and
$I\otimes\hat U(\phi)$ acts on the entire permuted vector.
Substitution into the equation of motion gives
\begin{align}
\frac{\mathrm{d}\widetilde V}{\mathrm{d}\phi}
=
-\widetilde Y_{\mathrm{eff}}(\phi)\widetilde V,
\label{eq:app_transformed_equation}
\end{align}
where
\begin{align}
\widetilde Y_{\mathrm{eff}}(\phi)
={}&
\left[I\otimes\hat U^\dagger(\phi)\right]
\hat Y_{\mathrm{eff}}(\phi)
\left[I\otimes\hat U(\phi)\right]
\nonumber\\
&+
I\otimes
\left[
\hat U^\dagger(\phi)
\frac{\mathrm{d}\hat U(\phi)}{\mathrm{d}\phi}
\right].
\label{eq:app_transformed_Yeff}
\end{align}
The transformed system-sector contribution and the rotating-frame
connection are, respectively,
\begin{align}
\overline Y_s
&=
\hat U^\dagger(\phi)
Y_s(\phi)
\hat U(\phi)
\nonumber\\
&=
-\frac{rs}{\alpha}\sigma_y
+\mathrm{i}\frac{\beta sc}{\alpha}\sigma_x
-\mathrm{i}\frac{\beta s^2}{\alpha}\sigma_z,
\nonumber\\
\Gamma
&=
\hat U^\dagger(\phi)
\frac{\mathrm{d}\hat U(\phi)}{\mathrm{d}\phi}
=
-\frac{\mathrm{i}}{2}\sigma_z.
\label{eq:app_rotation_connection}
\end{align}
The complete transformed generator therefore takes the block form
\begin{align}
\widetilde Y_{\mathrm{eff}}(\phi)
=
\begin{pmatrix}
\overline Y_s+q_{00}(\phi)I+\Gamma
&
q_{01}(\phi)I
\\
q_{10}(\phi)I
&
\overline Y_s+q_{11}(\phi)I+\Gamma
\end{pmatrix}.
\label{eq:app_final_Yeff}
\end{align}
Although $\overline Y_s$ and $\Gamma$ are constant, the coefficients
$q_{jk}(\phi)$ originate from $Y_a(\phi)$ and remain path dependent for a
generic non-equatorial constant-latitude loop. Its circuit realization
therefore requires time-dependent elements, which increases the complexity
of both construction and simulation. The implementation and validation of
such time-dependent circuits are beyond the scope of this work and are left
for future investigation.

\section{Monte Carlo Procedure for Static Resistor Tolerances}
\label{app3}

All simulations employ the full LTspice macro-model of the AD823A
operational amplifier, which includes its typical input offset voltage of
$0.2\,\mathrm{mV}$---more than three orders of magnitude below the
volt-scale node signals (see Table~\ref{tab:parameters})---together with
its finite gain-bandwidth product and slew-rate limits. The agreement
between simulation and theory in Figs.~\ref{fig:A} and
\ref{fig:phase_results} is obtained with this macro-model, indicating that
op-amp offsets are negligible at the present signal scale. Initial node
voltages are imposed on the integrator capacitors at $t=0$; the
$\sim1\,\mu\mathrm{s}$ settling transient mentioned in Sec.~\ref{sec3D} is
negligible compared to the evolution cycle and does not enter the
final-voltage readout.

For the $n$th Monte Carlo realization, the actual resistance of the $j$th
coupling resistor is modeled as
\begin{align}
R_j^{(n)}(r)
=
R_j(r)
\left[
1+\eta\xi_j^{(n)}
\right],
 \nonumber\\
\xi_j^{(n)}\sim\mathcal{U}(-1,1),
\quad
j=1,\ldots,16,
\label{eq:app_resistor_tolerance}
\end{align}
where $R_j(r)$ is the nominal resistance at purity $r$, $\eta$ is the
prescribed tolerance level, and $\xi_j^{(n)}$ is independently sampled for
every resistor and every realization. Equivalently, the relative resistance
deviation satisfies
\begin{align}
\delta_j^{(n)}
=
\eta\xi_j^{(n)}
\sim
\mathcal{U}(-\eta,\eta).
\label{eq:app_relative_deviation}
\end{align}
Thus, $\eta$ specifies the maximum magnitude of the uniformly distributed
relative deviation, rather than a one-standard-deviation component error.
The $16$ coupling resistors comprise, in the $A$ block, two resistors of
value $R_{A0}$ (realizing the $\pm h_{00}$ couplings), four of value
$R_\kappa$ (realizing the $\pm r/\alpha$ couplings), and two of value
$R_{A1}$ (realizing the $\pm h_{11}$ couplings), with $R_{B0}$ and $R_{B1}$
appearing analogously in the $B$ block; their $r$-dependent nominal values
are listed in Table~\ref{tab:parameters}. The gain-setting resistor pairs
of the eight unity-gain inverters are held at their nominal values in all
realizations (each pair consists of two equal resistors, so that the
nominal gain is exactly $-1$); a mismatch within a pair can be absorbed
into an effective tolerance of the corresponding coupling branch.

For each realization $n$, the set
$\left\{\xi_1^{(n)},\xi_2^{(n)},\ldots,\xi_{16}^{(n)}\right\}$
is generated once and held fixed during the complete scan over $r$. Although
the nominal values $R_j(r)$ vary with $r$, the relative deviations
$\delta_j^{(n)}$ remain unchanged. Each realization therefore represents one
fabricated nonideal circuit with fixed component values.

The same normalized random vectors $\xi_j^{(n)}$ are used for the
$\eta=0.5\%$ and $\eta=1.0\%$ datasets, with only the tolerance amplitude
changed. This common-random-number construction reduces incidental sampling
differences when comparing the two tolerance levels.

The purity is scanned on a uniform grid with spacing $\Delta r=0.002$ over
the range shown in Fig.~\ref{fig:phase_results}. For each circuit
realization, the LTspice scan yields a complex overlap
curve $I_n(r)
=
\operatorname{Re}I_n(r)
+
\mathrm{i}\operatorname{Im}I_n(r)$
with overlap amplitude
\begin{align}
|I_n(r)|
=
\sqrt{
\left[\operatorname{Re}I_n(r)\right]^2
+
\left[\operatorname{Im}I_n(r)\right]^2
}.
\label{eq:app_overlap_amplitude}
\end{align}
To refine the location of the overlap-amplitude minimum beyond the scan
resolution $\Delta r$, we identify the scan point at which
$|I_n(r)|^2$ is minimal and perform a local quadratic fit using this point
and its two nearest neighbors on either side (up to four in total). The
vertex of the fitted parabola is taken as the effective critical purity
$r_c^{(n)}$.

After $N$ Monte Carlo realizations, any unsuccessful LTspice runs or
invalid local fits would be excluded; in the present datasets all runs
converged and yielded valid fits, so that $N_{\mathrm{valid}}=N=100$ for
both tolerance levels. The mean extracted critical purity and its sample
standard deviation are
\begin{align}
\bar r_c
&=
\frac{1}{N_{\mathrm{valid}}}
\sum_{n=1}^{N_{\mathrm{valid}}}
r_c^{(n)},
\nonumber\\
\sigma_{r_c}
&=
\left[
\frac{1}{N_{\mathrm{valid}}-1}
\sum_{n=1}^{N_{\mathrm{valid}}}
\left(
r_c^{(n)}-\bar r_c
\right)^2
\right]^{1/2}.
\label{eq:app_std_rc}
\end{align}
The reported Monte Carlo result is $r_c^{\mathrm{MC}}
=\bar r_c\pm\sigma_{r_c}$, where $\sigma_{r_c}$ describes the circuit-to-circuit spread of the
extracted critical purity; it is not the standard error of the mean.

At each fixed value of $r$, the mean overlap amplitude is evaluated as
\begin{align}
\left\langle |I(r)|\right\rangle
=\frac{1}{N_{\mathrm{valid}}}
\sum_{n=1}^{N_{\mathrm{valid}}}
|I_n(r)|,
\label{eq:app_mean_absI}
\end{align}
and its sample standard deviation is
\begin{align}
\sigma_{|I|}(r)
=
\left[
\frac{1}{N_{\mathrm{valid}}-1}
\sum_{n=1}^{N_{\mathrm{valid}}}
\left(
|I_n(r)|
-
\left\langle |I(r)|\right\rangle
\right)^2
\right]^{1/2}.
\label{eq:app_std_absI}
\end{align}
The tolerance-limited resolvability used in the main text is then
\begin{align}
\mathrm{SNR}_{\mathrm{tol}}(r)
=
\frac{|I_0(r)|}{\sigma_{|I|}(r)},
\label{eq:app_snr_tol}
\end{align}
where $I_0(r)$ is the ideal overlap. The boundaries of the low-resolvability
interval are obtained by linearly interpolating between adjacent scan points
that bracket $\mathrm{SNR}_{\mathrm{tol}}(r)=1$.
\bibliography{Review,Review1,Review2}
\end{document}